\documentclass[12pt]{article}
\usepackage{amssymb,graphicx,bm,harvard,endnotes}
\renewcommand{\footnote}{\endnote}

\begin{document}
\title{Financial Probabilities from Fisher Information}
\author{Raymond J.~Hawkins$^{\dag}$ and B.~Roy Frieden$^{\ddag}$ \\
$^{\dag}$Mulsanne Capital Management, 220 Montgomery Street \\
Suite 506, San~Francisco, CA  94104, USA \\
$^{\ddag}$Optical Sciences Center, University of Arizona \\
Tucson, AZ 85721, USA}
\maketitle

\begin{abstract}
We present a novel synthesis of Fisher information and asset
pricing theory that yields a practical method for reconstructing
the probability density implicit in security prices.
The Fisher information approach to these inverse problems
transforms the search for a probability density into the solution
of a differential equation for which a
substantial collection of numerical methods exist.  We illustrate
the potential of this approach by calculating the probability
density implicit in both bond and option prices. Comparing the
results of this approach with those obtained using maximum entropy
we find that Fisher information usually results in probability densities
that are smoother than those obtained using maximum
entropy.
\end{abstract}


\section{Introduction}
Probability laws that derive from a variational principle provide an operational calculus for the incorporation of new knowledge.  This feature, long exploited in the physical sciences, is becoming increasingly popular in finance and economics where, for example, maximum entropy has found application as a useful and practical computational approach to financial economics~\cite{Maasoumi93,Sengupta93,Golanetal96,FombyHill97}.  It is generally felt that
a candidate probability law $p(x)$ should be as non-informative and smooth (in
some sense) as possible while maintaining consistency with the
known information about $x$. While this criterion has often been used to motivate the use of maximum entropy~\cite{BuckMacaulay90}, other variational approaches provide similar - and potentially superior - degrees of smoothness to probability laws~\cite{Frieden88,Edelman01} and it is the purpose of this paper to show that Fisher information~\cite{Frieden98} - which provides just such a variational approach - can be used to reconstruct probability densities of interest in financial economics.  In particular, Fisher information provides a variational calculus, a well developed computational approach to the estimation of probability laws and yields a probability law where smoothness is ensured across the range of support; in contrast to maximum entropy where smoothness tends to be concentrated in regions where the probability density is very small.

Since maximum entropy is comparatively well known in financial economics and as it shares with Fisher information a common variational structure, we shall use it as a point of comparison when we review Fisher information in Sec.~\ref{sec:theory}.  In that section we shall see that Fisher information yields not an analytic expression for the probability law, but rather, a differential equation for the probability law that is well known in the physical sciences:  the Schroedinger equation.  Using this fortunate correspondence we can exploit decades of development of computational approaches to the Schroedinger equation in the construction of probability laws in finance and economics.  

Like other variational approaches the Fisher information approach depends upon prior information in the form of equality constraints, and we shall explore this in Sec.~\ref{sec:examples} where examples of increasingly complex structure will be given.  In particular, exploiting the correspondence between (i) the resulting differential equations for probability laws in these examples from financial economics and (ii) the Schroedinger equation, will permit the generation both of yield curves from observed bond prices and of probability densities from observed option prices in new and efficient ways.  In addition, we shall see that the probability densities generated using Fisher information are, in general, smoother than those obtained from maximum entropy.  A brief summary of our results is given in Sec.~\ref{sec:summary} which is followed by two appendices where details of the calculations are presented in somewhat expanded form.
\section{Theory}
\label{sec:theory} 
When observed data are expectation values, their relationship to 
the probability density is that of a linear integral, and the 
calculation of the probability density from these observations is
a linear inverse problem.  

In the most general case we have $M$ observed
data values $d_1, \; \ldots , \; d_M = \left\{ d_m \right\}$ that are
known to be averages of known functions $\left\{ f_m(x) \right\}$ and 
are related to a probability density function $p(x)$ by
\begin{equation}
\int_a^b f_m(x) p(x) dx = d_m, \; m=1, \ldots, \; M \;  .\label{eq:lie}
\end{equation}
As there are an infinite number of probability densities that satisfy Eq.~\ref{eq:lie}, a regularizer is 
commonly introduced to impose a structural constraint on the characteristics
of $p(x)$ and to choose a particular density.  This can be accomplished by 
constructing a Lagrangian and employing the variational calculus.  In the
case of maximum entropy, the regularizer is the Shannon entropy~\cite{shannon48}\begin{equation}
H \equiv -\int_a^b p(x) \ln \left [ p(x) \right ] dx \; , \label{eq:shannon}
\end{equation}
with which one can form the Lagrangian
\begin{equation}
\int_a^b p(x) \ln \left [ p(x) \right ] dx + \sum_{m=1}^M \lambda_m \left [
\int_a^b f_m(x) p(x) dx - d_m \right ] \; ,
\end{equation}
where the $\{ \lambda_m \}$ are the Lagrange undetermined multipliers.  For this Lagrangian the extremum solution with zero first variation is known
to be~\cite{jaynes68}
\begin{equation}
p_{ME}(x)  =  \frac{1}{Z(\lambda_1, \; \ldots \;
\lambda_M )} \exp \left [ -\sum_{m=1}^M \lambda_m f_m(x) \right ] \; , \label{eq:pme1}
\end{equation}
where
\begin{equation}
Z(\lambda_1, \; \ldots \; \lambda_M )  =  \int_a^b \exp
\left [ -\sum_{m=1}^M \lambda_m f_m(x) \right ] dx \; . \label{eq:pme2}
\end{equation}
By comparison, in most uses of the Fisher information approach the Shannon entropy is simply replaced with Fisher information~\cite{fisher25} in its shift-invariant form
\begin{equation}
I = \int_a^b \frac{p'(x)^2}{p(x)}dx \; , \label{eq:fi}
\end{equation}
where the prime denotes differentiation with respect to $x$.  Consequently the Lagrangian
becomes
\begin{equation}
\int_a^b \frac{p'(x)^2}{p(x)}dx + \sum_{m=1}^M \lambda_m \left [
\int_a^b f_m(x) p(x) dx - d_m \right ] \; .
\end{equation}
As before we seek an extremum solution with zero first variation.  This time,
however, the Euler-Lagrange equations result in a differential equation for
$p(x)$:
\begin{equation}
2 \frac{d}{dx}  \left [ \frac{p'(x)}{p(x)} \right ] + \left [
\frac{p'(x)}{p(x)} \right ]^2 - \sum_{m=1}^M \lambda_m f_m(x) = 0
\; . \label{eq:elfi}
\end{equation}
This equation can be simplified through the use of a probability amplitude
function $q(x)$ where $q^2(x) \equiv p(x)$.  Substituting the probability
amplitude into Eq.~\ref{eq:elfi} yields the expression
\begin{equation}
\frac{d^2 q(x)}{dx^2} = \frac{q(x)}{4} \sum_{m=1}^M \lambda_m 
f_m(x) \; , \label{eq:swe}
\end{equation}
known in the physical sciences as the Schroedinger equation.  

Thus we see that, although the use of Fisher information results in a 
differential equation for $p(x)$ instead of an algebraic function, the 
resulting differential equation for $q(x)$ is one for which a
number of analytic solutions are known and for which a substantial collection
of numerical solutions exist.  Furthermore, as we shall see in the next 
section, the Fisher information solutions are, in general, smoother than those
obtained using maximum entropy.
\section{Examples}
In this section we use the Fisher information approach to motivate practical computational approaches to the extraction of probability densities and related quantities from financial observables.  In each case we contrast the results obtained by this method with those of maximum entropy.  We begin in Secs.~\ref{sec:fi} and~\ref{sec:perp} with examples from fixed income where observables are expressible in terms of partial integrals of a probability density and of first moments of the density.  In Secs.~\ref{sec:option} and~\ref{sec:vol} we finish the examples by examining the canonical problem of extracting risk-neutral densities from observed option prices and extend this analysis by showing how Fisher information can be used to calculate a generalized implied volatility.
\label{sec:examples}
\subsection{The Probability Density in the Term Structure of Interest Rates:  Yield Curve Construction}
\label{sec:fi} 
A basis for the use of information theory in
fixed-income analysis comes from the perspective developed by~\citeasnoun{BrodyHoughston01} who
observed that the price of a zero-coupon bond - also known as the
discount factor $D(t)$ - can be viewed as a
complementary probability distribution where the maturity date is
taken to be an abstract random variable.  The associated probability density $p(t)$ satisfies $p(t)>0$ for all $t$, and
\begin{eqnarray}
D(t) & = & \int_t^{\infty} p(\tau)d\tau \; , \label{eq:dfi} \\
 & = & \int_0^{\infty} \Theta(\tau-t) p(\tau)d\tau \; ,
\end{eqnarray}
where $\Theta(x)$ is the Heaviside step
function\footnote{$\Theta(x)=0$ if $x<0$ and 1 otherwise.}.

To illustrate the differences between the Fisher information and 
maximum entropy approaches to term structure estimation we consider 
the case of a single zero coupon bond of tenor $T = 10$ years and a 
price of 28 cents on the dollar or, equivalently, a discount factor
of 0.28~\footnote{These particular parameters were chosen so that the trigonometric functions resolve into simple fractions.}.  The maximum entropy solution, Eq.~\ref{eq:pme1}, is
\begin{equation}
p_{ME}(t) = \frac{e^{-\lambda \Theta (t - T)}}{T + 1/\lambda} \; , \label{eq:mets}
\end{equation}
and the corresponding discount factor is $D(T) = 1/(\lambda T + 1)$ from which the Lagrange multiplier is found to be 0.2571.

The Fisher information solution is obtained by solving Eq.~\ref{eq:swe} with the constraints of normalization and one discount factor:
 \begin{equation}
\frac{d^2 q(t)}{dt^2} = \frac{q(t)}{4} \left [ \lambda_0 + \lambda_1 \Theta (t-10) \right ] \; ,
\end{equation}
This problem is known\footnote{See, for example, Section 22 of~\cite{LandauLifshitzV377}.}  to have the solution
\begin{equation}
q(t) = 
\cases{
    A \cos ( \alpha t ) &if $t \leq T$;\cr
    B e^{-\beta t} &if $t > T$,
    }
\end{equation}
where the coefficients $A$ and $B$ are determined by requiring that $q(t)$ and $q'(t)$ 
(or the logarithmic derivative $q'(t)/q(t)$) match at $t=T$.  Matching
the logarithmic derivative yields
\begin{equation}
\tan(\alpha T) = \beta / \alpha \; .
\end{equation}
Choosing $\alpha T = \pi /4$ and matching amplitudes we find that $B = A \exp(\pi/4)/\sqrt{2}$,
\begin{equation}
p_{FI}(t) = 
\cases{
    \frac{ 8 \alpha \cos^2 ( \alpha t )}{\pi + 4} &if $t \leq T$;\cr
    \frac{ 8 \alpha e^{\pi/2 - \ln2 -2\beta t}}{\pi + 4} &if $t > T$;
    } \label{eq:fits}
\end{equation} 
and $D(T) = 2/(\pi+4) = 0.28$.  

\begin{figure}
\includegraphics{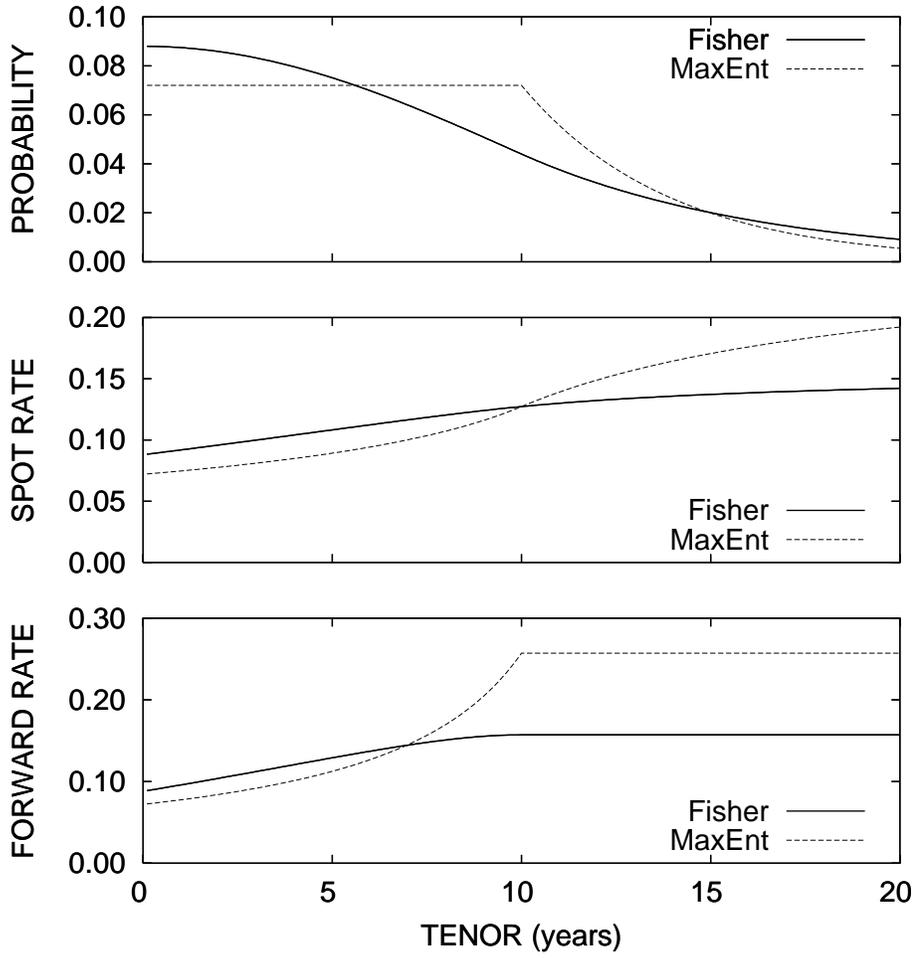}
\caption{The probability density and derived interest rates for a single ten year discount factor of 0.28} \label{fig:zcb}
\end{figure}

The maximum entropy and Fisher information probability densities (Eqs. \ref{eq:mets} and \ref{eq:fits} respectively) and the corresponding
term structure of interest rates are illustrated in Fig.~\ref{fig:zcb}.
In the uppermost panel we see both probability densities as a function 
of tenor with the Fisher information result denoted in all panels by 
the solid line.  As expected, the maximum entropy result is uniform until 
the first (and in this case only) observation is encountered; beyond which
a decaying exponential is observed.  The Fisher information result is smoother reflecting the need to match both the amplitude and derivative at the
data point.  

Looking at the probability density is difficult to make any real aesthetic choice between the
two results.  For this we turn to the two important derived quantities:  the 
spot rate $r(t)$ and the forward rate $f(t)$ that are related to the discount
factor by
\begin{eqnarray}
D(t) & = & e^{-r(t)t} \; , \\
 & = & e^{\int_0^t f(\tau) d\tau} \; .
\end{eqnarray}
The spot rates are shown in the middle panel of Fig.~\ref{fig:zcb}.  Both 
methods yield a smooth result with the Fisher information solution showing 
less structure than the maximum entropy solution.  A greater difference 
between the two methods is seen in the lowermost panel of Fig.~\ref{fig:zcb} where the forward rate is shown.  It is the structure of this function that is often looked to when assessing
the relative merits of a particular representation of the discount factor.  The
forward rate reflects the structure of the probability density as expected from
the relationship
\begin{equation}
f(t) = p(t)/D(t) \; ,
\end{equation}
with the maximum entropy result showing more structure than the Fisher 
information result due to the continuity at the level of the first derivative
imposed on $p(t)$ by the Fisher information approach.

It is a comparatively straightforward matter to extend this approach to the construction of a term structure that is consistent with any number of arbitrarily spaced zero coupon bonds.  A particularly convenient computational approach based on transfer matrices is presented in Appendix A. 

While previous work on inferring the term structure of interest rates from observed bond prices has usually focused on the somewhat {\em ad hoc} application of splines to the spot or forward rates~\cite{McCullough75,VasicekFong82,Shea85,Fisheretal95}, the work of~\citeasnoun{FrishlingYamamura96} that minimized $d f(t) / dt $ is similar in spirit to the Fisher information approach.  Their paper dealt with the often unacceptable results that a straightforward application of splines to this problem of inference can produce.  In some sense, Fisher information can be seen as an information-theoretic approach to imposing the structure sought by Frishling and Yamamura on the term structure of interest rates.\footnote{A similar pairing on the minimization of $d^2 f(t) / dt^2$ is seen in the work of~\citeasnoun{AdamsVanDeventer94} and the recent entropy work of~\citeasnoun{Edelman01}.}  
\subsection{The Probability Density in a Perpetual Annuity}
\label{sec:perp}
Material differences between the probability densities generated
by Fisher information and maximum entropy are also seen when the observed
data are moments of the density - a common situation in financial
applications.  The value of a perpetual annuity $\xi$, for example, is given by~\cite{BrodyHoughston01,BrodyHoughston02}
\begin{equation}
\xi = \int_0^{+\infty} \tau p(\tau) d\tau \; . \label{eq:perp}
\end{equation}
This first-moment constraint provides an interesting point of
comparison for the maximum entropy approach employed by Brody and Houghston and our Fisher
information approach.  The maximum entropy solution is known to be
\begin{equation}
p_{ME} (t) = \frac{1}{\xi} \exp \left ( -t / \xi \right ) \;
\end{equation}
while the Fisher information solution is known~\cite{Frieden88} to be
\begin{equation}
p_{FI} (t) = c_1 {\rm Ai}^2 \left ( c_2 t \right ) \; ,
\end{equation}
where ${\rm Ai}(x)$ is Airy's function and the constants $c_i$ are
determined uniquely by normalization and by the constraint of
Eq.~\ref{eq:perp}.
\begin{figure}
\includegraphics{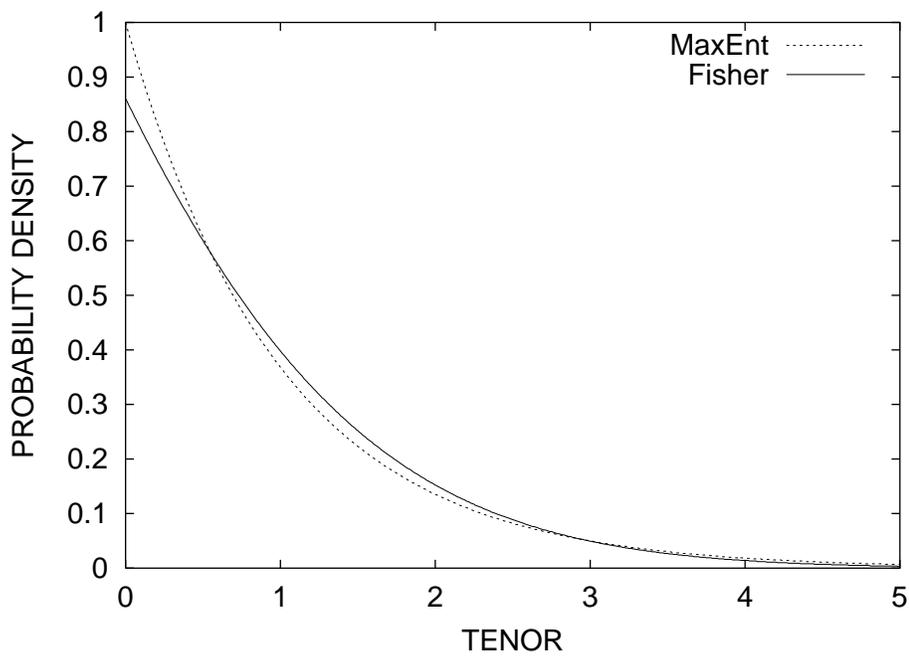}
\caption{The probability density functions associated with a
perpetual annuity with a price of 1.0.} \label{fig:perp}
\end{figure}
We show these two probability densities as a function of tenor for
a perpetual annuity with a price of \$1.00 in Fig.~\ref{fig:perp}.  The two solutions are qualitatively similar in appearance; both monotonically decrease with tenor and are quite smooth.  However, since the Fisher information
solution starts out at zero tenor with a lower value than the maximum entropy solution and
then crosses the maximum entropy solution so as to fall off more slowly than maximum entropy solution in
the mid-range region, the Fisher information solution seems to be even smoother.  
In the context of Fisher information, a measure of smoothness is
the size of the Cramer-Rao bound (CRB) 1/$I$~\cite{Frieden88} that, for the Fisher information
solution is found to be 1.308 by integrating Eq.~\ref{eq:fi}.  The
CRB for the maximum entropy solution is 1.000.  Hence, by the criterion of
maximum Cramer-Rao bound, the Fisher information solution is significantly
smoother.  We also compared the relative smoothness of these
solutions using Shannon's entropy, $H$, (Eq.~\ref{eq:shannon}) as
a measure.  The maximum entropy solution is found to have an $H$ of 1.0, while
the Fisher information solution has an $H$ of 0.993.  Hence the maximum entropy solution does
indeed have a larger Shannon entropy than does the Fisher information solution,
but it is certainly not much larger.  Since the two solutions
differ much more in their CRB values than in their Shannon $H$
values, it appears that the CRB is a more sensitive measure of
smoothness, and hence biasedness, of a probability law.

An interesting result emerges from an examination of the
smoothness as a function of range.  Since over the
range $0 \leq T \leq 5$ shown in Fig.~\ref{fig:perp} it appears that the Fisher information ought to
be measurably smoother than the maximum entropy solution by any criterion, we
also computed the Shannon $H$ values over this interval alone. The
results are $H = 0.959$ for maximum entropy and $H = 0.978$ for Fisher information:  the Fisher information
solution has the larger entropy.  This shows that the Fisher information solution
is smoother than the maximum entropy solution, even by the maximum entropy criterion, over
this {\em finite} interval.  It follows that the maximum entropy solution has the
larger entropy overall only due to its action in the long tail
region $T > 5$.  Note, however, that in this region the maximum entropy
solution would have negligibly small values for most purposes.

Hence, the indications are that the criterion of maximum entropy places an
unduly strong weight on the behavior of the probability density
in the tail regions.  Note by comparison that over
the interval $0 \leq T \leq 5$ the CRB values for the
two solutions still strongly differ, with 1.317 for the Fisher information
solution and 1.007 for the maximum entropy.  Moreover, these are close to their
values as previously computed over the entire range $0 \leq T \leq
\infty$.  Hence, Fisher information gives comparatively lower weight to the tail
regions of the probability densities.
\subsection{The Probability Density in Option Prices}
\label{sec:option}  While the reconstruction of probability
densities from observed option prices has been the topic of much
research (c.f.~\cite{Jackwerth99} and references therein), the expectation of global smoothness
in the resulting probability densities has often not been achieved
through the use of techniques that focus globally (e.g. maximum entropy~\cite{Stutzer94,Hawkinsetal96,BuchenKelley96,Stutzer96,Hawkins97,Avellaneda98}).  As with term-structure estimation, various
approaches to smoothing probability densities have been often introduced
in an {\em ad hoc} manner. Fisher information, however,
provides a natural manner for the introduction of a measure of
local smoothness into the option-based density reconstruction
problem.

For purposes of illustration, let us consider the expression for
the price of a European call option $c(k)$:
\begin{equation}
c(k) = e^{-r t} \int_{0}^{\infty} \max(x-k,0) p(x) dx \; ,
\label{eq:call}
\end{equation}
where $k$ is the strike price, $t$ is the time to expiration, and $r$ is the risk-free interest rate.  We take $c(k)$, $k$, $r$, and $t$ to be known and ask what $p(x)$ is associated with the observed $c(k)$.  The observed call
value $c(k)$ can also be viewed as the mean value of the function
$e^{-r t} \max(x-k,0)$, and it is from this viewpoint that a
natural connection between option pricing theory and information theory can be made.  

Given a set of $M$ observed call prices $\{c(k_m)\}$ the Fisher Information solution is obtained by solving Eq.~\ref{eq:swe} subject to these constraints and normalization,
\begin{equation}
\frac{d^2 q(x)}{dx^2} = \frac{q(x)}{4} \left [ \lambda_0 + \sum_{m=1}^M \lambda_m
\max(x-k_m,0) \right ] \; . \label{eq:swe1}
\end{equation}
While the numerical method described in Appendix A can be applied to Eq.~\ref{eq:swe1} by replacing the terms multiplying $q(t)$ on the right-hand side of Eq.~\ref{eq:swe1} with a stepwise approximation or by treating the solution between strike prices as a linear combination of Airy functions, Eq.~\ref{eq:swe1} can be integrated directly using the Numerov method (cf.~Appendix B).\footnote{In this paper we focus solely on the equilibrium solutions:  those densities that correspond to the lowest value of $\lambda_0$.  Equation~\ref{eq:swe1} is solved by a collection of $\{\lambda_0, q(x)\}$ pairs; the solutions $q(x)$ corresponding to other $\lambda_0$ being nonequilibrium solutions.  These solutions are the subject of a forthcoming paper.}  The maximum entropy solution can be obtained in a straightforward manner substituting Eq.~\ref{eq:pme1} into Eq.~\ref{eq:call} and varying the $\{\lambda_m\}$ to reproduce the observed option prices.

\begin{figure}
\includegraphics{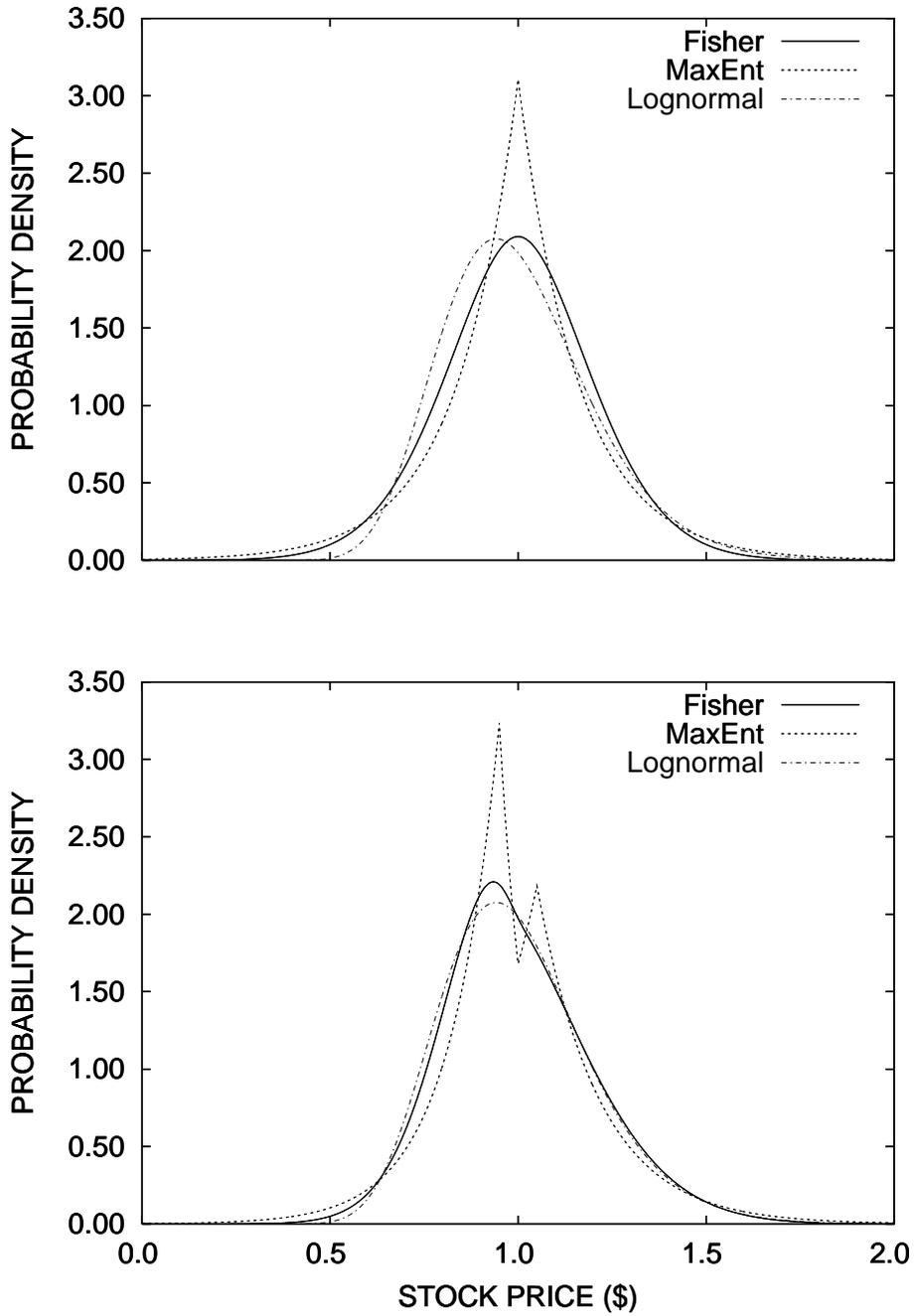}
\caption{Probability density functions associated with
option prices.} \label{fig:option}
\end{figure}

We applied the Fisher information approach to the example discussed in~\citeasnoun{Edelman01} of 3 call options on a stock with an
annualized volatility $\sigma$ of 20\% and a time to expiration
$t$ of 1 year.  In this example the interest rate $r$ is set to
zero, the stock price $S_t$ is \$1.00 and the
three option prices with strike price $k$ of \$0.95, \$1.00, and
\$1.05 have Black-Scholes prices \$0.105,
\$0.080, and \$0.059 respectively. We note that the
stock can also be viewed in this example as a call option with a
strike price of \$0.00.  Given these `observations' together
with the normalization requirement on $p(x)$ we generated the solutions shown in Fig.~\ref{fig:option}.  The smooth dash-dot curve is the lognormal distribution with which the `observed' option prices were sampled\footnote{The lognormal distribution is 
$$
p(S_T, T | S_t, t) = \frac{1}{S_T \sqrt{2 \pi \sigma^2 (T-t)}}
\exp \left ( -\frac{\left [ \ln(S_T/S_t) - r (T-t)\right ]^2}{2
\sigma^2 (T-t)} \right )
$$
from which the Black-Scholes call price
\begin{eqnarray*}
c(k) & = & S_t N(d_1) - k e^{-r(T-t)} N(d2) \,  \\
d_1 & \equiv & \frac{\ln (S_t/k) + (r + \sigma^2 / 2)(T-t)}{\sigma
\sqrt{T-t}} \\
d_2 & \equiv & d_1 - \sigma \sqrt{T-t}
\end{eqnarray*}
can be derived.}, and is the distribution we are trying to recover from the option prices.  The Fisher information result is given by the solid curve and the MaxEnt result is the dashed curve that was calibrated to the observed option prices.  

We began by considering the case of the probability density implicit in a {\em single} at-the-money (i.e.~$k = \$1.00$) option price and the results are shown in the upper panel.  The maximum entropy solution is a sharply peaked product of two exponentials.  In contrast the Fisher information solution has, by virtue of requiring continuity of the first derivative, a much smoother appearance and is much closer in appearance to the lognormal density from which the option price was generated.  We do, however, see that with only a single option price the asymmetric features of the lognormal density are not recovered.  

Adding the option prices at the $k=\$0.95$ and $k=\$1.05$ (from Edelman example) strikes to the problem results in the densities shown in the lower panel of Fig.~\ref{fig:option}.  With this information the agreement between the Fisher information solution and the lognormal density is improved substantially with the peak of the density functions nearly the same and the characteristic asymmetry of the lognormal density now seen in the Fisher information result..  The maximum entropy solution continues to be sharply peaked and to deviate substantially from the lognormal density.  Sharply peaked densities have appeared in previous work involving limited financial observables (see, for example,~\cite{KuwaharaMarsh94}) and this feature should not be interpreted as a specific indictment of the maximum entropy approach.  In previous work~\cite{Hawkinsetal96} we generated smooth implied densities from S\&P-500 index options by assuming a lognormal prior based on the at-the-money-option ($k=1$) option instead of the uniform prior assumed in Eq.~\ref{eq:shannon} and forming a Lagrangian using the Kullback-Leibler entropy~\cite{Kullback59}
\begin{equation}
G = -\int_a^b p(x) \ln \left [ \frac{p(x)}{r(x)} \right ] dx \; ,
\end{equation}
where $r(x)$ is the prior - in our case lognormal - density.  This points to one of the key differences between the maximum entropy and Fisher information approached: needed assumptions.  When observations are limited - as they often are in financial economics applications - smooth maximum entropy densities often require a prior density:  Fisher information simply imposes a continuity constraint.
\subsection{A Measure of Volatility}
\label{sec:vol}
In addition to providing a way of calculating the density implied by option prices, Fisher information yields, via the Cramer-Rao bound, a simultaneous measurement of the intrinsic uncertainty in this system.  This uncertainty principle is, perhaps, most easily seen if we recall that Fisher information is generally presented in terms of a system specified by a parameter $\theta$ with a probability density $p(x|\theta)$ where
\begin{equation}
\int_a^b p(x|\theta) dx = 1\; ,
\end{equation}
and the Fisher information is
\begin{equation}
\int_a^b p(x|\theta) \left [ \frac{\partial p(x|\theta)/\partial \theta}{p(x|\theta)}\right ]^2 dx \; .
\end{equation}
For distributions that can be parameterized by a location parameter $\theta$, Fisher information can then be defined with respect to the family of densities  $\{p(x-\theta)\}$ (cf.~\cite{CoverThomas91}).  This transformation leads directly to Eq.~\ref{eq:fi} for the Fisher information.  It also implies that the mean-square error $e^2$ in the estimate of that location parameter is related to this Fisher information, $I$, by the Cramer-Rao inequality $e^2 I \geq 1$.  If, for example, $p(x)$ is a Gaussian, the minimum Cramer-Rao bound is $I = 1/\sigma^2$ with $\sigma^2$ being the variance.  Given that the probability density reconstructed from option prices is the conditional density of underlying asset levels at option expiration and given that a set of option prices is a measurement of the probability density it follows that the Fisher information, via the Cramer-Rao bound, provides a measure of the mean-square error in the estimate of the level of the underlying asset at expiration provided by the option prices and, thus, a natural generalization of the concept of implied volatility to non-Gaussian underlying-asset distributions.
\section{Discussion and Summary}
\label{sec:summary} 
Fisher information provides a variational approach to the estimation of probability laws that are consistent with financial observables.  Using this approach one can employ well developed computational approaches from formally identical problems in the physical sciences.  The resulting probability laws posses a degree of smoothness consistent with priors concerning the probability density and/or quantities derived therefrom.  In this paper we have shown how Fisher information can be used to solve some canonical inverse problems in financial economics and that the resulting probability densities are generally smoother than those obtained by other methods such as maximum entropy.  Fisher information also has the virtue of providing a natural measure of, and computation approach for determining, implied volatility via the Cramer-Rao inequality.

Comparing the maximum entropy approach to that of Fisher information it was seen that the former has the virtue of simplicity, always being an exponential.  The Fisher information solution is always a differential equation for the probability density.  Although somewhat more complicated, this is, in some sense, a virtue as the probability densities of physics and, presumably of finance and economics, generally obey differential equations.  Indeed, it is in this application that we see the greatest potential for future applications of Fisher information in finance and economics.  For example, in addition to the problems examined in Sec.~\ref{sec:examples}, recent developments in macroeconomic modeling have focused upon placing all or nearly all heterogeneous microeconomic agents within one stochastic and dynamic framework.  This has the effect of replacing the replicator dynamics of evolutionary games or the Malthusian dynamics of Friedman with the backward Chapman-Kolmogorov equation~\cite{Aoki98,Aoki01}.  Such a statistical mechanical representation of macroeconomics provides a natural link to Fisher information.  The solutions of the Schroedinger equation (Eq.~\ref{eq:swe}) are known to be stationary solutions to the Fokker-Plank equation~\cite{Gardiner96,Risken96}.  More generally, however, the entire Legendre-transform structure of thermodynamics can be expressed using Fisher information in place of the Shannon entropy - an essential ingredient for constructing a statistical mechanics~\cite{Friedenetal99} - and it has been shown~\cite{Friedenetal02} that both equilibrium and non-equilibrium statistical mechanics can be obtained from Eq.~\ref{eq:swe}:  the output of a constrained process that extremizes Fisher information.  

The elegant and powerful Lagrangian approach to physics, in use for over 200 years, has until a recent application of the Fisher information approach~\cite{FriedenSoffer95,Frieden98} not had a formal method for constructing Lagrangians.  This formalism for deriving Lagrangians and probability laws that are consistent with observations should prove to be a very useful tool in finance and economics.
\section*{Acknowledgments}
We thank Prof.~David Edelman for providing a preprint of his work and for stimulating discussions.  We also thank Dana Hobson, Prof.~David A.B.~Miller, Leif Wennerberg, and Prof.~Ewan Wright for helpful discussions and suggestions.

\renewcommand{\theequation}{A-\arabic{equation}}
\setcounter{equation}{0}  
\section*{Appendix A}  
To generalize the amplitude-matching that we used in Sec.~\ref{sec:fi} we use a transfer matrix method that is employed often for formally similar problems in quantum electronics~\cite{Yehetal77}\footnote{Our presentation of this derivation was derived, in part, from unpublished optoelectronics notes of Prof.~David A.B.~Miller.}.  Given a set of discount factors $\{D(T_m)\}$ we seek a solution to 
\begin{equation}
\frac{d^2 q(t)}{dt^2} = \frac{q(t)}{4} \left [ \lambda_0 + \sum_{m=1}^M \lambda_m
\Theta(t-T_m,0) \right ] \; .
\end{equation}
Between tenors, say $T_{m-1}$ and $T_m$ it is straightforward to show that the probability amplitude $q(t)$ is given by
\begin{equation}
q(t) = A_m e^{i \omega_m (t-T_m)} + B_m e^{-i \omega_m (t-T_m)} \; ,
\end{equation}
where $A_m$ and $B_m$ are coefficients to be determined by the matching conditions, $i = \sqrt{-1}$, and 
\begin{equation}
\omega_m = \frac{1}{2}\sqrt{\sum_{j=0}^m \lambda_j} \; ,
\end{equation}
with $\{\lambda_j\}$ being the Lagrange undetermined multipliers.  We now consider the match across the $m^{th}$ tenor.  Since $t - T_m = 0$, continuity of $q(t)$ gives
\begin{equation}
q(T_M) = A_L + B_L = A_{m+1} + B_{m+1} \; ,
\end{equation}
where the subscript $L$ denotes the value of the coefficients of $q(t)$ immediately to the left of $T_m$.  Continuity of the first derivative of $q(t)$ gives
\begin{equation}
A_L - B_L = \Delta_m \left (A_{m+1} + B_{m+1} \right ) \; ,
\end{equation}
where $\Delta_m = \omega_{m+1}/\omega_m$.  These two equations combine to give
\begin{equation}
\left [ 
\begin{array}{c}
A_L \\ B_L
\end{array}
\right ] = \frac{1}{2}
\left [
\begin{array}{cc}
1 + \Delta_m & 1 - \Delta_m \\
1 - \Delta_m & 1 + \Delta_m
\end{array}
\right ]
\left [
\begin{array}{c}
A_{m+1} \\ B_{m+1}
\end{array}
\right ]
\end{equation}
Since $A_m = A_L \exp (-i \omega_m \tau_m)$ and $B_m = A_L \exp (i \omega_m \tau_m)$ where $\tau_m = T_{m+1} - T_m $ it follows that the coefficients for $q(t)$ in layer $m$ are related to those in layer $m+1$ by
\begin{equation}
\left [ 
\begin{array}{c}
A_m \\ B_m
\end{array}
\right ] = \frac{1}{2}
\left [
\begin{array}{cc}
e^{-i \omega_m \tau_m} & 0 \\
0 & e^{i \omega_m \tau_m}
\end{array}
\right ]
\left [
\begin{array}{cc}
1 + \Delta_m & 1 - \Delta_m \\
1 - \Delta_m & 1 + \Delta_m
\end{array}
\right ]
\left [
\begin{array}{c}
A_{m+1} \\ B_{m+1}
\end{array}
\right ]
\end{equation}
Given this relationship we can generate the function $q(t)$ sequentially, beginning with the region beyond the longest tenor that we shall take to be $T_{m+1}$.  In this region we require that the solution be a decaying exponential $\exp (-\omega_{m+1} t)$ which implies that $B_{m+1} = 0$ and that $\omega_{m+1}$ be a positive real number.  Given a $\{\lambda_m\}$ and setting  $A_{m+1} = 1$ , it is now a straightforward matter to calculate the fixed income functions described in Sec.~\ref{sec:fi}.  This transfer matrix procedure maps $\{\lambda_m\}$ into a discount function and, in a straightforward manner, into coupon bond prices.
\renewcommand{\theequation}{B-\arabic{equation}}
\setcounter{equation}{0}  
\section*{Appendix B}  
Our choice of the Numerov method to solve Eq.~\ref{eq:swe1} follows from the popularity of this method for solving this equation in quantum mechanics~\cite{Koonin86}.  In this appendix we present a brief overview of the method and it's application to the option problem discussed in Section~\ref{sec:option}.

The Numerov method provides a numerical solution to\footnote{The Numerov method can, in fact, be used to solve numerically the somewhat more general equation
$$
\frac{d^2 q(x)}{d x^2} + G(x)q(x) = S(x) \;.
$$
The development of the algorithm follows the same line as that illustrated in this Appendix.}
\begin{equation}
\frac{d^2 q(x)}{d x^2} + G(x)q(x) = 0 \; ,
\label{eq:genode}
\end{equation}
that begins by expanding $q(x)$ in a Taylor series
\begin{equation}
q(x \pm \Delta x) = \sum_{n=0}^{\infty} \frac{(\pm \Delta x)^n}{n!} \frac{d^n q(x)}{dx^n} \; .
\end{equation}
Combining these two equations and retaining terms to fourth order yields
\begin{equation}
\frac{q(x+\Delta x) + q(x-\Delta x) - 2q(x)}{(\Delta x)^2} =
\frac{d^2 q(x)}{dx^2} + \frac{(\Delta x)^2}{12} \frac{d^4 q(x)}{dx^4} + \mathcal{O}((\Delta x)^4) \; ,
\label{eq:basis}
\end{equation}
The second derivative term in Eq.~\ref{eq:basis} can be cleared by substituting Eq.~\ref{eq:genode} and the fourth derivative term can be cleared by differentiating Eq.~\ref{eq:genode} two more times
\begin{eqnarray}
\frac{d^4 q(x)}{dx^4} & = & -\frac{d^2 \left ( G(x) q(x) \right )}{dx^2} \; ,\\
 & \cong & -\frac{G(x+\Delta x)q(x+\Delta x)+G(x-\Delta x)q(x-\Delta x)-2G(x)q(x)}{(\Delta x)^2} \; ,\nonumber
\label{eq:fourth}
\end{eqnarray}
and substituting this result into Eq.~\ref{eq:genode}.  Performing these substitutions and rearranging terms we arrive at the Numerov algorithm:
\begin{equation}
q_{i+1} = \frac{2 \left ( 1 - \frac{5(\Delta x)^2}{12} G_n \right ) }
{\left ( 1 + \frac{(\Delta x)^2}{12} G_{n+1} \right ) } q_n - 
\frac{\left ( 1 + \frac{(\Delta x)^2}{12} G_{n-1} \right ) }
{\left ( 1 + \frac{(\Delta x)^2}{12} G_{n+1} \right ) } q_{n+1} + \mathcal{O}((\Delta x)^6) \; .
\end{equation}
To apply this to the option problem it is useful to note that the calculation can be separated into two stages:  first, the calculation of the $\{\lambda_0,q(x)\}$ pair that solves the eigenvalue problem for a fixed set of $\{\lambda_m\}_{m>0}$, and second, the search for the set $\{\lambda_m\}_{m>0}$ for which $q(x)$ of the corresponding $\{\lambda_0,q(x)\}$ pair reproduces the observed option prices.

The $\{\lambda_0, q(x)\}$ pair for a given $\{\lambda_m\}_{m>0}$ is obtained by integrating $q(x)$ from both the lower and upper limits of the $x$ range and varying $\lambda_0$ until the logarithmic derivatives of each solution match at an arbitrary mid point.  In this example that point was $x=1$ and $\lambda_0$ was found with a 1-dimensional minimization algorithm using the difference in the logarithmic derivatives as a penalty function and the minimum of the function   $\sum_{m=1}^M \lambda_m \max(x-k_m,0)/4$ as the initial guess for $\lambda_0$.

Once the logarithmic derivatives are matched the option prices can be calculated and compared with the observed option prices.  The results of this comparison form the penalty function for a multidimensional variation of the $\{\lambda_m\}_{m>0}$ to obtain agreement between the observed and calculated option prices.

\theendnotes \bigskip

\bibliographystyle{agsm}
\bibliography{fisher}
\end{document}